\documentclass[a4paper,fleqn]{cas-dc}

\usepackage[authoryear,longnamesfirst]{natbib}
\usepackage{graphicx}
\usepackage{pgfplots}
\usepackage{pgf-pie}
\pgfplotsset{width=7cm,compat=1.5}
\usepackage{multirow}
\usepackage{tikz}
\definecolor{source}{gray}{0.95}
\definecolor{highlight}{gray}{0.9}
\definecolor{bblue}{HTML}{4F81BD}
\definecolor{rred}{HTML}{C0504D}
\definecolor{ggreen}{HTML}{9BBB59}
\definecolor{ppurple}{HTML}{9F4C7C}

\def\tsc#1{\csdef{#1}{\textsc{\lowercase{#1}}\xspace}}
\tsc{WGM}
\tsc{QE}
\tsc{EP}
\tsc{PMS}
\tsc{BEC}
\tsc{DE}
\usepackage{xspace}
\usepackage{graphicx}
\usepackage{ifthen}
\usepackage[normalem]{ulem} 
\usepackage{xcolor,colortbl}
\usepackage{hyperref}

\newboolean{showedits}
\setboolean{showedits}{true} 
\ifthenelse{\boolean{showedits}}
{
	\newcommand{\del}[1]{\textcolor{red}{\sout{#1}}} 
	\newcommand{\nbe}[3]{
		{\colorbox{#3}{\bfseries\sffamily\scriptsize\textcolor{white}{#1}}}
		{\textcolor{#3}{\sf\small$\blacktriangleright$\textit{#2}$\blacktriangleleft$}}}
}{
	\newcommand{\del}[1]{} 
	
	\newcommand{\nbe}[3]{}
}

\newboolean{showcomments}
\setboolean{showcomments}{true} 
\newcommand{\id}[1]{$-$Id: scgPaper.tex 32478 2010-04-29 09:11:32Z oscar $-$}

\ifthenelse{\boolean{showcomments}}
 {
 	\newcommand{\nbc}[3]{
 		{\colorbox{#3}{\bfseries\sffamily\scriptsize\textcolor{white}{#1}}}
		{\textcolor{#3}{\sf\small$\blacktriangleright$\textit{#2}$\blacktriangleleft$}}}
	
 }{
 	\newcommand{\nbc}[3]{}
 	
 }

\usepackage[most]{tcolorbox}
\ifthenelse{\boolean{showedits}}
{
  \newtcolorbox{inserted}{%
       title=Inserted text:,
       colframe=blue,colback=blue!5!white,
       breakable,
       leftrule=0mm, 
       bottomrule=0mm,
       rightrule=0mm,
       toprule=0mm,
       arc=0mm, outer arc=0mm,
       oversize
  }
  \newtcolorbox{deleted}{%
       title=Deleted text:,
       colframe=red,colback=red!5!white,
       breakable,
       leftrule=0mm, 
       bottomrule=0mm,
       rightrule=0mm,
       toprule=0mm,
       arc=0mm, outer arc=0mm,
       oversize
  }
  \newtcolorbox{refactored}{%
       title=Rewritten text:,
       colframe=blue,colback=red!5!white,
       breakable,
       leftrule=0mm, 
       bottomrule=0mm,
       rightrule=0mm,
       toprule=0mm,
       arc=0mm, outer arc=0mm,
       oversize
  }
}{

}
\newboolean{isblinded}
\setboolean{isblinded}{true}
\ifthenelse{\boolean{isblinded}}
{\newcommand\blind[1]{BLINDED\xspace}}
{\newcommand\blind[1]{#1\xspace}}

\newcommand{\commented}[1]{}

\newcommand{\eg}{\emph{e.g.,}\xspace}
\newcommand{\ie}{\emph{i.e.,}\xspace}

\usepackage{tcolorbox}

\newcommand{\boxit}[2][gray!15]{%
    \begin{tcolorbox}[
        colback=#1,    
        colframe=black, 
        width=8.3cm,    
        arc=0mm,        
        boxrule=0.5pt,  
        fontupper=\small,
        left=2pt,       
        right=2pt,      
    ]
        \emph{#2}
    \end{tcolorbox}
}

\definecolor{source}{gray}{0.9}

\begin{document}

\title [mode = title]{
Time to Separate from StackOverflow and Match with ChatGPT for Encryption
}    

\shorttitle{
Time to Separate from StackOverflow and Match with ChatGPT for Encryption
}

\author{Ehsan Firouzi}[orcid=0009-0000-7563-4196]
\ead{ehsan.firouzi@tu-clausthal.de}

\author{Mohammad Ghafari}[orcid=0000-0002-1986-9668]
\ead{mohammad.ghafari@tu-clausthal.de}

\address{Technische Univertität Clausthal, Germany}

\shortauthors{E. Firouzi and M. Ghafari}

 \lstset{
        language=Java,
         basicstyle=\ttfamily\footnotesize,
	keywordstyle=\color{blue}\bfseries,
        stringstyle=\color{DarkViolet},
	mathescape=true,
	showstringspaces=false,
	keepspaces=true,
	numbers=left,                    
        numbersep=4pt,                  
	breakautoindent=true,
	backgroundcolor=\color{source},
	upquote=true, 
	columns=fullflexible} 
 \newcommand{\lct}[1]{{\textsf{\textup{#1}}}}
\lstnewenvironment{codesnippet}{%
	\lstset{%
		frame=single,
		framerule=0pt,
		mathescape=false
	}
}{}

\begin{abstract}
Cryptography is known as a challenging topic for developers.
We studied StackOverflow posts to identify the problems that developers encounter when using Java Cryptography Architecture (JCA) for symmetric encryption. 
We investigated security risks that are disseminated in these posts, and we examined whether ChatGPT helps avoid cryptography issues.
We found that developers frequently struggle with key and IV generations, as well as padding. 
Security is a top concern among developers, but security issues are pervasive in code snippets.
ChatGPT can effectively aid developers when they engage with it properly. Nevertheless, it does not substitute human expertise, and developers should remain alert.
\end{abstract}



\maketitle
\section{Introduction}

The analysis of security issue reports across numerous open-source projects on GitHub revealed a concerning trend: the proliferation of security issues is on the rise, while their resolution progresses slowly and only a small group of developers are involved in the process~\citep{Buhlmann2022}.
Despite the crucial role of Cryptography in the seamless integration of security into our digital world, developers struggle with existing cryptography libraries.
These libraries often do not support common operations, lack sufficient abstraction, and have poor documentation quality~\citep{mindermann2018, hazhirpasand2021_b, patnaik2019}. 
Hence, API misuses are likely and so does the presence of security vulnerabilities.
For instance, the analysis of cryptography in 489 open-source Java projects has revealed that 85\% included API misuses~\citep{hazhirpasand2020}.
These issues are also present in proprietary software systems. Notably, researchers have identified weak encryption algorithms and legacy encryption modes in critical infrastructure~\citep{Wetzels2023}.

Java Cryptography Architecture (JCA) is the most widely adopted cryptography API, and symmetric encryption is the foremost adopted cryptography operation in software systems.  
Most of the top 100 cryptography questions on StackOverflow, sorted by views and score, are about symmetric encryption.
Similarly, it is adopted in 64\% of the top 100 GitHub projects, sorted by stars, that use JCA~\citep{nadi2016}.

Unlike prior studies that are broad, 
in this paper, we focused specifically on symmetric encryption with JCA, providing a detailed view on its challenges for developers.
We blended qualitative and quantitative analyses to uncover developer issues and common security risks observed on the StackOverflow website, and we evaluated the effectiveness of ChatGPT in addressing these issues.
In particular, we investigated the following three research questions:

\textbf{RQ$_1$:} What are common developer challenges in symmetric encryption?

We manually inspected 400 StackOverflow posts that are about JCA symmetric encryption and found that the majority of reported problems (\ie 214 posts) are at the ``cipher object initialization'' stage. 
The primary challenges were in key and initialization vector (IV) management. 
There were 116 posts that highlighted issues encountered during the ``cipher object instantiation'' stage, particularly related to padding, encryption mode, and algorithm selection. 
We identified 99 posts that reported problems during the ``encryption/decryption'' stages, with encoding being the most prevalent issue. 
The ``transmission of parameters (key/IV)'' had the lowest number of reported issues (\ie 21 posts).
They were mainly linked to encrypting keys using various algorithms and facing difficulties related to keystores.
We also found a total of 128 exceptions, with ``BadPaddingException'' and
``InvalidKeyException'' is the most common, occurring 106 times in the posts.
Nevertheless,  
only 45 exceptions (\ie 42\%) were directly related to padding and key issues, making it challenging for developers to pinpoint the root causes.

\textbf{RQ$_2$:} What are the security risks present in the shared JCA code on StackOverflow?

The examination of 13 security rules in 400 posts revealed a striking number of 327 posts (\ie 82\%) with security violations.
The use of ECB and CBC encryption modes were the most common violations followed by hard-coded keys. 
We collected the symptoms of security violations and searched for them across all 3,426 symmetric encryption posts on StackOverflow.
The findings revealed 5,305 violations in 3,174 posts, averaging 1.7 violations in 92\% of StackOverflow posts.
In general, we observed a lack of adherence to best practices, such as those for password-based key generation, but the adoption of 256-bit keys for AES and GCM encryption mode have increased in recent years.

\textbf{RQ$_3$:} How effective is ChatGPT for addressing developer issues in symmetric encryption?

We provided 100 StackOverflow questions to ChatGPT (GPT-3.5) and recorded the answers.
We observed that 
it provided a ``working'' solution. 
Nevertheless, 
they are problematic from a security perspective.
Precisely,
when we provided the exact StackOverflow questions to ChatGPT, it transferred almost every violation from the question to its answer.
When we explicitly prompted to provide a ``secure solution'', it cleared violations in 42 questions.
ChatGPT detected DES as an insecure algorithm and ECB as a vulnerable encryption mode almost in every case.
We could clear violations in 26 more questions when we pinpointed the exact line where a violation existed.
In the end, it adopted a weak key generation function in two questions and included a CBC mode in 30 questions, which ChatGPT did not flag as a violation even for client-server scenarios.

In summary,
the contributions of this paper are the following.

\begin{itemize}
    
\item 
We provided a comprehensive study on Java developer challenges in symmetric encryption. 

\item
We discovered that security violations are prevalent in the code examples shared on the StackOverflow website, threatening novices who blindly copy and paste code into their programs.

\item 
We highlighted the potential of ChatGPT in complementing human expertise and demonstrated its superiority in evaluating code security on a line-by-line basis.

\item We shared our dataset of symmetric encryption posts, along with
the detailed results of our manual investigations, the security expression rules, as well as the links to ChatGPT responses.\footnote{https://github.com/Ehsan-Firouzi/SymmetricEncryptionChallenges}

\end{itemize}

The rest of this article is organized as follows.  
We describe our research methodology in Section~\ref{sec:DataCollection}.
We present our results in Section~\ref{sec:result} and discuss our findings in Section~\ref{sec:Discussion}.
We explain threats to the validity of our study in Section~\ref{sec:ThreatstoValidity} and make an overview of related work in Section~\ref{sec:relatedwork}.
Finally, we conclude this paper in Section~\ref{sec:Conclusion}.


\section{Study Setup}
\label{sec:DataCollection}

We relied on the Stack Exchange Data Dump released on March 8, 2023~\citep{ArchStackexchange}.
We extracted the StackOverflow posts that are related to symmetric encryption and randomly selected a representative subset of these posts for manual inspection.  

\subsection{Data Collection}

The \texttt{Cipher} class in JCA offers encryption and decryption functionalities.
It supports both block and stream cipher symmetric encryption algorithms. 
In short, stream ciphers convert one symbol of plaintext directly into a symbol of ciphertext, whereas block ciphers, which are more modern, encrypt a group of plaintext symbols as one block.
Table~\ref{JCAAlgo} lists these algorithms. 

\begin{table}[htbp]
 \centering
  \caption {Symmetric algorithms in JCA}
\begin{tabular}{p{3.8cm} p{3.5cm}}
\hline
\textbf{Symmetric Encryption} & \textbf{Algorithms}                                                                   \\ \hline
\rowcolor[HTML]{EFEFEF} 
Block Cipher          & \begin{tabular}[c]{@{}l@{}}AES\\ DES\\DESede(3DES)\\ Blowfish\end{tabular} \\
Stream Cipher         & \begin{tabular}[c]{@{}l@{}}RC2, RC4, RC6\\ ChaCha20\end{tabular}             \\ \hline
\end{tabular}
\label{JCAAlgo}
\end{table}

\emph{Full dataset.}
We searched the StackOverflow for cipher instances instantiated with a symmetric encryption algorithm. 
We utilized the regular expression (RegEx) shown in Listing~\ref{cipherpatternJCA} to search within both questions and accepted answers. We discarded posts where the patterns were only found within other answers, as viewers of the posts pay greater attention to the aforementioned sections. This filtering process resulted in 3,426 posts.

\begin{lstlisting}[label=cipherpatternJCA, caption=
RegEx for capturing symmetric encryption uses in JCA,language=C,basicstyle=\footnotesize, numbers=none]
Cipher\.getInstance\(("|\&quot)(AES|DES|DESede|RC|Blowfish|ChaCha20)
\end{lstlisting}

\emph{Sample dataset.}
We chose a subset of posts for a manual inspection, and to ensure that the findings represent the entire dataset, we included 400 posts.
This sample size yields a 95\% confidence level with a less than 5\% (4.5\%) margin of error.
We sought to incline our sample data towards recent posts and those with high scores. Therefore,
we selected 40\% of the posts between 2020 and 2023, representing the most recent posts, 30\% from the most popular ones (highest scores), and the other 30\% completely random. 

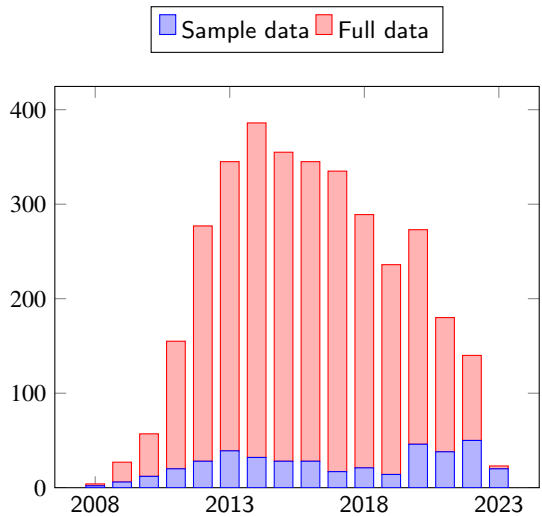
\begin{figure}
\centering    
\begin{tikzpicture}
\begin{axis}[
  width  = 0.95*\linewidth,
    ybar stacked,
    ymin=0,
    bar width=7pt,
    legend style={at={(0.5,1.2)},
      anchor=north,legend columns=-1
      },
    symbolic x coords={
2008,
2009,
2010,
2011,
2012,
2013,
2014,
2015,
2016,
2017,
2018,
2019,
2020,
2021,
2022,
2023
},
    ]
\addplot+[ybar] plot coordinates {
(2008,2)
(2009,6)
(2010,12)
(2011,20)
(2012,28)
(2013,39)
(2014,32)
(2015,28)
(2016,28)
(2017,17)
(2018,21)
(2019,14)
(2020,46)
(2021,38)
(2022,50)
(2023,20)
 };
\addplot+[ybar] plot coordinates {
 (2008,2)
(2009,21)
(2010,45)
(2011,135)
(2012,249)
(2013,306)
(2014,354)
(2015,327)
(2016,317)
(2017,318)
(2018,268)
(2019,222)
(2020,227)
(2021,142)
(2022,90)
(2023,3)
 };
\legend{\strut Sample data, \strut Full data}
\end{axis}
\end{tikzpicture}

\caption{Distribution of symmetric encryption posts}
\label{fig:distposts}
\end{figure}

In summary, we gathered a total of 3,426 posts for our ``full dataset'' and chose 400 posts to form our ``sample dataset''.
Figure~\ref{fig:distposts} illustrates the distribution of these posts in each year, and Table~\ref{tab:DatasetDetails} lists further details about our data.

\begin{table}[htbp]
 \centering
  \caption {Further details about the collected StackOverflow posts}
\begin{tabular}{p{1cm} p{1.2cm} p{1.2cm} p{1cm} p{1cm}}

\hline
            \textbf{Dataset} & \textbf{\#Solved Posts} & \textbf{\#Pending Posts} & \textbf{Score AVG.} & \textbf{View AVG.} \\ \hline
            \rowcolor[HTML]{EFEFEF} 
Full    & 1,723         & 1,703          & 2          & 3,263     \\ 
Sample  & 251          & 149           & 5          & 13,892    \\ \hline
\end{tabular}
\label{tab:DatasetDetails}
\end{table}
 
\begin{figure*}
\includegraphics[width=\textwidth]{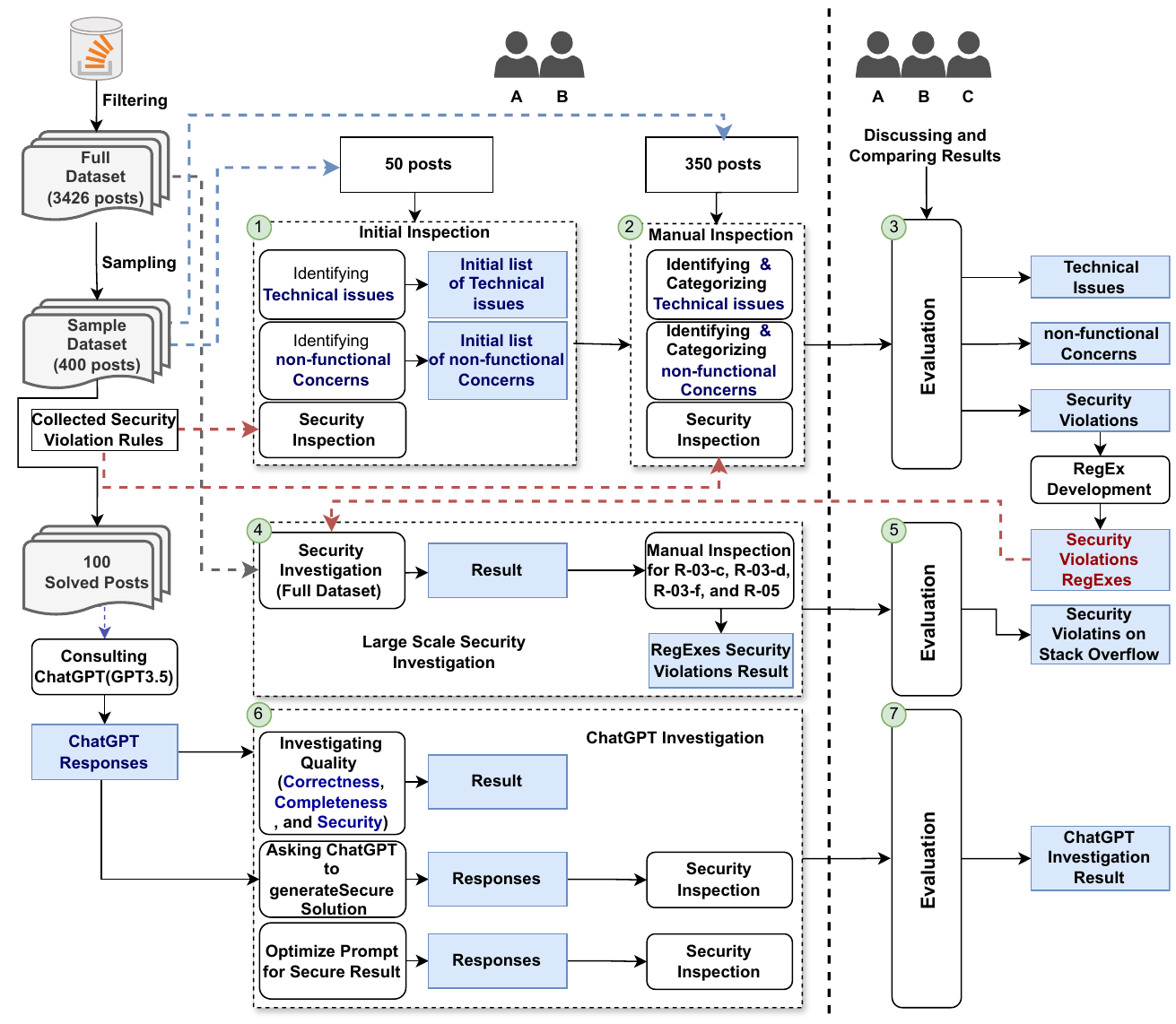}
\caption{Overview of our research methodology}
\label{fig:Method}
\end{figure*}

\subsection{Methodology}

Figure~\ref{fig:Method} shows the overview of our methodology.
Three individuals (A1, A2, and A3) were involved, each possessing practical knowledge in cryptography and over three years of Java programming experience. 

\subsubsection{Developer Challenges}
\label{Sec:DevChalengeMethod}
To identify and categorize the technical issues (\ie, root causes of the problems), and identify developers' non-functional concerns, we conducted a lightweight open-coding-like process. This process involved three phases as follows:

\emph{Phase I (Collecting an Initial List of Technical Issues and Concerns).} 
A1 and A2 reviewed 50 posts (questions, their accepted answers, and comments) together.
For each post, guided by official sources~\citep{javaReferenceGuide, ferguson2011cryptography, chung2012non}, they identified and recorded the technical issues (i.e., root causes of the problems), the stage at which these issues occurred, and non-functional concerns that were explicitly raised.
This step resulted in the compilation of two initial lists, including (1) technical issues and (2) non-functional concerns.

\emph{Phase II (Categorization).}
A1 and A2 independently reviewed the remaining 350 posts to complete the identification and categorization of technical issues and non-functional concerns. 

\emph{Phase III (Evaluation).}
A1 and A2 compared their findings and categorizations.
They discussed any disagreements until a consensus was reached. In instances where differing opinions persisted for certain posts, A3, who had not yet reviewed those posts, was consulted. 
Ultimately, the results were finalized using a majority voting mechanism (Cohen's k = 0.86).

\subsubsection{Security Analysis}
\label{method:Security Analysis}

We followed three phases to uncover security risks. We conducted them  
simultaneously during the investigation of developer challenges.

\emph{Phase I (Collecting Security Violation Rules).}
Recognizing the challenge posed by the incomplete nature of code snippets on StackOverflow, making it difficult to assess their security using available tools, we opted for a meticulous manual approach. We consulted the state-of-the-art in crypto misuse analysis and compiled a list of security violation rules for symmetric encryption. These rules, detailed in Table~\ref{table:SecRules}, were primarily sourced from tools such as CryLogger~\citep{piccolboni2021} and CogniCrypt~\citep{kruger2019}, as well as insights from a recent study in this domain~\citep{AutoDetecJavaCrypto}.

\emph {Phase II (Security Violations Investigation).} A1 and A2 separately examined the security of each post according to rules collected in Phase I. In particular, for each post in the sample dataset, they checked whether any of the violation rules were present in the question or accepted answer, and collected the code snippets associated with every violation.
 
 \emph {Phase III (Evaluation).}
 A1 and A2 compared their
results. They discussed any disagreements
until a consensus was reached. In instances where differing
opinions persisted for certain posts, A3 was consulted. Ultimately, the
results were finalized using a majority voting mechanism
(Cohen’s k = 0.86). 
At the end of this phase, through the study of insecure code snippets, we developed a set of regular expressions to enable us to detect security violations among code snippets. Recognizing the challenge of automatically resolving values passed as parameters in functions, often overlooked by static analysis tools~\citep{AfroseDetectionTools}, we sought ways to enhance precision. To address this, we manualized a part of our investigation, specifically for rules R-03-c, R-03-d, R-03-f, and R-05, and we considered this concern in the regular expressions (RegExes).
 
\emph{Phase IV (Large Scale Security Analysis).} To acquire overall insights into the state of security in the StackOverflow posts:
 
 \begin{enumerate}
     \item We applied the regular expressions extracted in the previous phase to the full dataset (3426 posts) for a comprehensive security analysis of all StackOverflow posts.
     \item Following this, A1 and A2 separately conducted manual checks on the results of the above-mentioned rules (in Section Phase III of Section~\ref{method:Security Analysis}) to identify insecure code snippets that violated these specified rules.
     \item Similar to phase III in section~\ref{Sec:DevChalengeMethod} A1, A2, and A3 evaluated the results (Cohen’s k = 0.91). The code snippets identified by RegExes for rules R-01, R-02-a, R-02-b, and R-03-g do not require additional manual checking, as they are confirmed to be 100\% vulnerable.   
 \end{enumerate}

\subsubsection{ChatGPT Analysis}

We consulted ChatGPT (GPT-3.5) to evaluate its potential in assisting developers with cryptography-related queries.
We randomly selected 100 posts with accepted answers from our pool of 400 sample posts (Evaluating the correctness and completeness of generated answers for questions without accepted answers is challenging because there is no answer to compare with the generated answers. This makes the study challenging and introduces a higher risk of human bias).
Each chosen post \emph{included at least one security violation}. 

\begin{enumerate}
    \item Firstly, we presented each ``exact'' question from the posts to ChatGPT, followed by the command ``answer this StackOverflow question and provide a code example''.' We collected its responses.
    \item A1 and A2 independently and manually:\\
    2.1.Examined the ChatGPT answers from a security perspective (by the help of collected security violation rules in Table~\ref{table:SecRules}) \\
    2.2. They evaluated the correctness and completeness of the generated answers. In this regard, for each question they compared the ChatGPT answer with the accepted answer and assessed if the ChatGPT answer was correct or misleading (Correctness) and if the answer fully addressed the question(Completeness).  as there is no accepted answer to compare with, making the study more challenging and with a higher risk of human bias.

    \item like phase III in section~\ref{Sec:DevChalengeMethod} , A1 and A2 compared results, resolving disagreements for consensus. A3 was consulted for persistent differences. Final results were determined by majority voting (Cohen's k = 0.82).
   \item Secondly, we explicitly instructed ChatGPT to generate a ``secure'' code example. and checked the answers from a security perspective.

  \item Lastly, for posts where ChatGPT's response remained insecure, we searched how we can optimize the prompt to get a more secure answer.

\end{enumerate}

\begin{table*}[t]
  \caption {Security violation rules applicable to symmetric encryption}
\begin{tabular}{lllll}
\hline
\textbf{Stage}                                                                                  & \textbf{Rule-ID}                                                      & \textbf{CWE-ID}                                                                                                                                                & \textbf{Violation}                                                                                                                       & \textbf{JCA API}                                               \\ 
\hline
\multirow{3}{*}{\begin{tabular}[c]{@{}l@{}}Cipher \\ Instantiation\end{tabular}}       & R-01                                                         & CWE-327                                                                                                                                               & Using weak  Algorithm                                                                                                           & Cipher                                                \\
                                                                                       & {\cellcolor[rgb]{0.902,0.902,0.906}}R-02-a                   & {\cellcolor[rgb]{0.902,0.902,0.906}}CWE-327                                                                                                           & {\cellcolor[rgb]{0.902,0.902,0.906}}Using ECB encryption mode                                                                   & {\cellcolor[rgb]{0.902,0.902,0.906}}Cipher            \\
                                                                                       & R-02-b                                                       & CWE-327                                                                                                                                               & Using CBC encryption mode                                                                                                       & Cipher                                                \\ 
\hline
\multirow{10}{*}{\begin{tabular}[c]{@{}l@{}}Cipher \\ Initialization-Key\end{tabular}} & {\cellcolor[rgb]{0.902,0.902,0.906}}R-03-a                   & {\cellcolor[rgb]{0.902,0.902,0.906}}CWE-798                                                                                                           & {\cellcolor[rgb]{0.902,0.902,0.906}}Using static or constant key                                                                & {\cellcolor[rgb]{0.902,0.902,0.906}}SecretKeyspec     \\
                                                                                       & \multirow{2}{*}{R-03-b}                                      & \multirow{2}{*}{CWE-330}                                                                                                                              & \multirow{2}{*}{Using static    salt for key derivation}                                                                        & PBEKeySpec                                            \\
                                                                                       &                                                              &                                                                                                                                                       &                                                                                                                                 & PBEParameterSpec                                      \\
                                                                                       & {\cellcolor[rgb]{0.902,0.902,0.906}}                         & {\cellcolor[rgb]{0.902,0.902,0.906}}                                                                                                                  & {\cellcolor[rgb]{0.902,0.902,0.906}}                                                                                            & {\cellcolor[rgb]{0.902,0.902,0.906}}PBEKeySpec        \\
                                                                                       & \multirow{-2}{*}{{\cellcolor[rgb]{0.902,0.902,0.906}}R-03-c} & \multirow{-2}{*}{{\cellcolor[rgb]{0.902,0.902,0.906}}\begin{tabular}[c]{@{}>{\cellcolor[rgb]{0.902,0.902,0.906}}l@{}}CWE-326 \\ CWE-330\end{tabular}} & \multirow{-2}{*}{{\cellcolor[rgb]{0.902,0.902,0.906}}Using salt 64 bits for key derivation.}                                    & {\cellcolor[rgb]{0.902,0.902,0.906}}PBEParameterSpec  \\
                                                                                       & \multirow{2}{*}{R-03-d}                                      & \multirow{2}{*}{\begin{tabular}[c]{@{}l@{}}CWE-326\\ CWE-330\end{tabular}}                                                                            & \multirow{2}{*}{Using iterations1000  for key derivation}                                                                       & PBEKeySpec                                            \\
                                                                                       &                                                              &                                                                                                                                                       &                                                                                                                                 & PBEParameterSpec                                      \\
                                                                                       & {\cellcolor[rgb]{0.902,0.902,0.906}}R-03-e                   & {\cellcolor[rgb]{0.902,0.902,0.906}}CWE-259                                                                                                           & {\cellcolor[rgb]{0.902,0.902,0.906}}Using hard-coded password                                                                   & {\cellcolor[rgb]{0.902,0.902,0.906}}PBEKeySpec        \\
                                                                                       & R-03-f                                                       & CWE-330                                                                                                                                               & Using   weak Random function for generating secret key                                                                          & KeyGenerator                                          \\
                                                                                       & {\cellcolor[rgb]{0.902,0.902,0.906}}R-03-g                   & {\cellcolor[rgb]{0.902,0.902,0.906}}CWE-327                                                                                                           & {\cellcolor[rgb]{0.902,0.902,0.906}}Using weak algorithms for generating secret key                                             & {\cellcolor[rgb]{0.902,0.902,0.906}}SecretKeyFactory  \\ 
\hline
\multirow{2}{*}{\begin{tabular}[c]{@{}l@{}}Cipher\\ Initialization-IV\end{tabular}}    & R-04-a                                                       & CWE-330                                                                                                                                               & Using static IV                                                                                                                 & IvParameterSpec                                       \\
                                                                                       & {\cellcolor[rgb]{0.902,0.902,0.906}}R-04-b                   & {\cellcolor[rgb]{0.902,0.902,0.906}}CWE-330                                                                                                           & {\cellcolor[rgb]{0.902,0.902,0.906}}Using a “badly-derived” Initialization Vector (IV)                                           & {\cellcolor[rgb]{0.902,0.902,0.906}}IvParameterSpec   \\ 
\hline
\begin{tabular}[c]{@{}l@{}}Parameters\\Transmission\end{tabular}                     & R-05                                                         & CWE-798                                                                                                                                               & \begin{tabular}[c]{@{}l@{}}Loading a keystore using an input stream \\ with a constant and non-null password value\end{tabular} & KeyStore                                             
\end{tabular}
\label{table:SecRules}
\end{table*}

\section{Result}
\label{sec:result}

We present our findings about developer challenges.
Then, we report our security investigation result, and lastly, we show to what extent ChatGPT can clear these issues. 

\subsection{Developer Challenges}
\label{subsec:RQ1}

We present technical issues as well as developer concerns discussed in 400 sample posts. 

\subsubsection{Technical Issues}
We found a total of 450 technical issues in 400 posts.
We categorized these issues into nine categories across four steps. 
The top common issues were related to the initialization step followed by issues in the instantiation step.
Table~\ref{Functional-Issues} lists the most common issues in each encryption stage.
We present these issues and share common root causes that we observed.

\begin{table}[]
\caption{Distribution of technical issues per stage}
\begin{tabular}{llcc}
\hline
                                  & \textbf{Problem}               & \textbf{\#Issues} & \textbf{\#Posts}               \\ \hline
\multirow{3}{*}{Instantiation}         & {\cellcolor[rgb]{0.902,0.902,0.906}}Padding                  &{\cellcolor[rgb]{0.902,0.902,0.906}} 89      & \multirow{3}{*}{116} \\ 
                                       & Encryption mode          & 49      &                      \\ 
                                       &{\cellcolor[rgb]{0.902,0.902,0.906}}EncryptionAlgorithm      &{\cellcolor[rgb]{0.902,0.902,0.906}} 15      &                      \\ \hline
\multirow{2}{*}{Initialization}        & Key                      & 121     & \multirow{2}{*}{214} \\ 
                                       & {\cellcolor[rgb]{0.902,0.902,0.906}}IV                       & {\cellcolor[rgb]{0.902,0.902,0.906}}106     &                      \\ \hline
\multirow{2}{*}{\begin{tabular}[c]{@{}l@{}}Encryption/\\ Decryption\end{tabular} } & Encoding                 & 54      & \multirow{2}{*}{99}  \\ 
                                       & {\cellcolor[rgb]{0.902,0.902,0.906}}Update/doFinal           & {\cellcolor[rgb]{0.902,0.902,0.906}}18      &                      \\ \hline
\multirow{2}{*}{Transmission} & \begin{tabular}[c]{@{}l@{}}Key\\ wrapping/Exchange\end{tabular}  & 12      & \multirow{2}{*}{21}  \\ 
                                       &{\cellcolor[rgb]{0.902,0.902,0.906}}Keystore                 & {\cellcolor[rgb]{0.902,0.902,0.906}}5       &                      \\ \hline
\end{tabular}
\label{Functional-Issues}
\end{table}

\emph{Cipher Instantiation.} 
This stage is for creating a Cipher object in Java for encryption and decryption operations using a factory method.
We discovered a total of 116 posts that were related to cipher instantiation.
The most common root causes were padding, encryption mode, and encryption algorithm.
In particular, 89 issues were associated with padding problems such as choosing an improper padding mode for a specific scenario, using different padding modes for encryption and decryption, and not explicitly specifying the padding mode. 
We identified 49 posts that had issues related to encryption mode.
These issues included the absence of an explicitly specified cipher mode, problems arising from dependencies between encryption mode and initialization vector (IV), and choosing different encryption modes for encryption and decryption.
Finally, we found 15 general posts about encryption algorithms such as how to adopt a specific algorithm.

\emph{Cipher Initialization.} 
The Cipher object has to be initialized with the appropriate encryption mode, key, and any necessary parameters such as initialization vectors.
Developers struggle the most in the cipher initialization step.
Key and vector initialization were the two most common root causes in 214 posts.
In particular, 121 posts were related to key initialization problems such as invalid key sizes (length), password-based key derivation problems, and using different keys for encryption and decryption.
There were 106 posts that dealt with issues related to IV. 
We observed common issues such as how to generate IVs, the dependency of IVs on encryption modes, using different IVs for encryption and decryption, and considerations regarding the size of IVs.

\emph{Encryption/Decryption.} 
The actual process of encryption or decryption occurs in this step, utilizing methods such as update and doFinal, along with encoding techniques.
We found that one-fourth of the posts discuss 
issues related to the encryption/decryption phase.
There were 54 posts about encoding issues, especially due to the use of different encoding systems during encryption and decryption.
In 18 posts, developers could not determine when to use \texttt{update()} and \texttt{doFinal()} methods.\footnote{The former method is to process data in chunks, whereas the latter method is used for the final step of encryption or decryption operations}

\emph{Transmission.} 
We require the encryption parameters, such as the encryption key, to decrypt a ciphertext.
The number of posts that discussed transmission issues was the lowest (\ie 21 posts).
The common issues that we observed were related to key wrapping and key exchange problems (\ie utilizing RSA and DH algorithms for exchanging symmetric keys ) in 12 posts,  
and keystore-related issues in five posts.


\subsubsection{Non-functional concerns} 

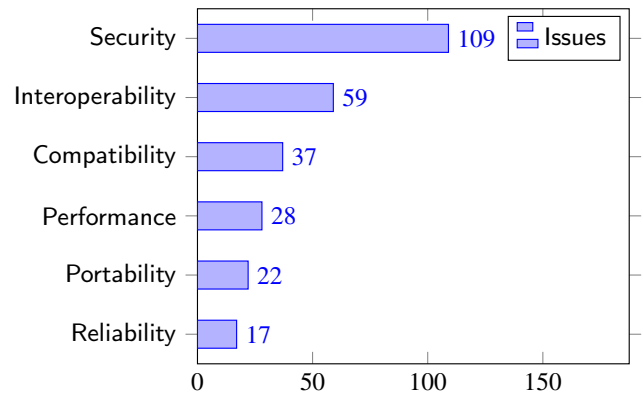
\begin{figure}
\centering

\resizebox{\columnwidth}{!}{%
\begin{tikzpicture}
 
\begin{axis} [xbar = .05cm,
    bar width = 10pt,
    xmin = 0, 
    xmax = 150, 
    enlarge x limits = {value = .25, upper},
    symbolic y coords={ Reliability, Portability, Performance,  Compatibility,Interoperability, Security},
    nodes near coords,
    ytick=data 
]
 
\addplot coordinates {   (17,Reliability)(22,Portability)(28,Performance) (37,Compatibility)(59,Interoperability)(109,Security)};
 \legend {Issues}
\end{axis}
 
\end{tikzpicture}}
\caption{Distribution of non-functional issues}
\label{Requierment Issues}
\end{figure}

We identified 272 non-functional concerns in 225 (out of 400) posts and categorized them into six main categories listed below.
The most recurring concerns were about security and interoperability.
Figure~\ref{Requierment Issues} shows a breakdown of the result.

\begin{itemize}
    
    \item Security: 
    We assigned posts to this category if the user explicitly inquired about security or relevant hints were present in the discussions.

    \item Performance:
    Posts in this category were concerned about enhancing the efficiency of the code.
    These posts mainly discussed issues related to optimizing memory usage, managing resources effectively, and improving the execution speed of the code (aiming for faster and more responsive performance).

    \item Reliability:
    This category of posts aimed to develop code that is stable and reliable, minimizing crashes~\citep{jalote2004measuring} and ensuring proper thread management.

    \item Compatibility:
    These posts deal with compatibility-related challenges such as issues arising from the code's incompatibility with specific providers or Java versions.

    \item Portability:
    Posts that discuss the ability of the code to be seamlessly transferred across different environments or platforms. 
    
    \item Interoperability:
    The posts within this category aimed to facilitate effective interaction and functionality between the code and code developed in other programming languages.
     
\end{itemize}

\emph{Security.}
The foremost concern, discussed in 109 posts, was about security.
The primary concern, highlighted in 46 posts, revolved around ensuring the unpredictability of initialization vectors (IV).
The second most prevalent concern pertained to the safety of encryption modes, mentioned in 32 posts.
We identified 29 posts expressing worries about the secure derivation of encryption keys from passwords.
Furthermore, 6 posts raised doubts about the security of chosen encryption algorithms.

\emph{Interoperability.}
We encountered 59 posts in which developers expressed concerns about interoperability.
There were 9 posts addressing issues with PHP, 9 posts related to Node.js, 8 posts highlighted challenges with C\#, and other posts (\ie 33) were related to other programming languages.
We observed that the underlying causes of interoperability issues were primarily related to differences in default encoding and padding methods.

\emph{Compatibility.}
Developers discussed compatibility in 37 posts.
The most common issues (\ie 14) were related to the JDK version.
We identified 5 issues about the absence of native support for PKCS7 padding, 4 issues due to Android versions, and 3 related to different service providers.

\emph{Performance.}
We found 28 posts that included performance concerns, of which 11 posts were about memory concerns, and 19 posts discussed speed concerns. 
Memory matters were about OutOfMemoryException for large files, which needed the use of CipherInputStream and CipherOutputStream instead of \texttt{cipher.update()} and \texttt{cipher.doFinal()}.
Speed matters were due to the buffer size or a lack of  \texttt{CipherInputStream} and \texttt{CipherOutputStream}.

\emph{Portability.}
Developers were concerned about the portability of their code in 22 posts. 
The discussions unanimously revolved around the default values of parameters on providers when developers do not explicitly specify them. 

\emph{Reliability.}
We found 17 posts that included reliability concerns, mainly related to unexpected behavior or crashes.
The most common issue (discussed in 5 posts) was about Cipher not being thread safe.

\boxit[yellow!20]{
Developers struggle a lot with key and IV generations, and padding.
Security and interoperability are their top concerns.  
}

\subsection {Security Risks}
\label{subsec:RQ2}

We investigated the presence of security issues in our sample posts.
We then developed a set of regular expressions that helped us to find such issues in the full dataset.

\begin{table}[]
\centering
\hspace{1cm}
\caption {Security violations in sample posts}
\begin{tabular}{llll}
\hline
 \textbf{Rule-ID}       &  \textbf{\#Solved Posts} &  \textbf{\#Pending Posts} &  \textbf{\#Total} \\ \hline
R-01 &  14 & 20 &  34\\ 
\rowcolor[HTML]{EFEFEF} R-02-a &  86  & 41 & 127 \\ 
R-02-b & 106  & 58  & 164 \\ 
\rowcolor[HTML]{EFEFEF} R-03-a &  73 & 27& 100  \\ 
R-03-b &  12 & 8 & 20\\
\rowcolor[HTML]{EFEFEF} R-03-c &  0 & 0 &  0\\ 
R-03-d &  4 & 0 & 4 \\ 
\rowcolor[HTML]{EFEFEF} R-03-e &  17 & 8 & 25 \\ 
R-03-f &  1 &1  &2 \\ 
\rowcolor[HTML]{EFEFEF} R-03-g &  18 & 15 & 33 \\
R-04-a & 36 & 12 &48  \\ 
\rowcolor[HTML]{EFEFEF} R-04-b &1 & 0 & 1\\
R-05 & 0  & 1& 1\\ \hline
\end{tabular}
\label{table:SecRulesviolAA}
\end{table}

\subsubsection{Security Rules} 
We found a total of 13 security rules in the literature that concern symmetric encryption.
Table~\ref{table:SecRules} lists these rules, indicates what exact APIs are involved, and provides CWE links.
The items highlighted in orange text do not necessarily indicate the insecure pattern; instead, they represent bad practices that are not secure enough.

\emph{R-01.}  
DES, RC2, and RC4 are outdated algorithms and have been compromised by modern cryptanalysis methods.
While 3DES is not broken but it was deprecated by NIST~\citep{SP800-67rev2}.
Blowfish is generally regarded as a strong and secure algorithm, but it has a 64-bit block size, making this algorithm susceptible to birthday attacks, especially in HTTPS context~\citep{Sweet32}.
Moreover, the GnuPG project recommended against using Blowfish for encrypting files larger than 4 GB due to its small block size~\citep{GNUPrivacyGuard}.
Blowfish implementations employ 16 rounds of encryption and are not vulnerable to known-plaintext attacks on weak keys, but a reduced-round variant of Blowfish is known to be vulnerable. 
Therefore, 
experts recommend to use AES~\citep{barker2020recommendation}, a symmetric encryption algorithm with a larger block size and stronger security features.

\emph{R-02.}
AES is widely considered as a secure symmetric encryption algorithm, but the choice of encryption mode is crucial for achieving maximum security. 
ECB (Electronic Codebook) mode is not secure when there is more than one block to encrypt,\footnote{Using AES without specifying an encryption mode is functionally equivalent to utilizing AES in ECB mode.} and CBC (Cipher Block Chaining) mode 
is secure for non client-server scenarios~\citep{Serge2002CBC}.
Indeed, CBC mode is not recommended for client-server scenarios as it requires careful IV management, proper padding, secure key management, and separate mechanisms for data integrity and authentication.
Due to our limited knowledge of the specific context in which code snippets are utilized, we consider any encryption performed with ECB or CBC modes as a security risk\footnote{CBC is secure insofar as it is designed to resist chosen-plaintext attacks, which it effectively does. However, CBC does not defend against chosen-ciphertext attacks, making it a bad practice for client/server scenarios.}.

\emph{R-03.}  
To ensure a robust initialization of the cipher-key, it is crucial to adhere to the guidelines outlined in R-03-a to R-03-g. 
The use of static or constant keys and salts for key derivation is not permitted at all (R-03-a and R-03-b). 
Instead, it is necessary to employ at least 64 bits of salt (R-03-c) and a minimum of 1000 iterations for key derivation (R-03-d). 
Hard-coded passwords (R-03-e) as well as weak random functions (R-03-f) or weak algorithms (R-03-g) for generating secret keys must be avoided, 

\emph{R-04.}
Proper initialization of an Initialization Vector (IV) is crucial for ensuring the security of AES encryption. 
To this end, R-04-a advises against using a static IV that may be predictable and void its purpose. 
There has to be a dynamic seed and proper randomness for IV generation (R-04-b) to ensure that each IV is unique and unpredictable for every encryption operation.

\emph{R-05.}
When loading a keystore, it is important to ensure that the password is retrieved from a secure external source, such as a database or a file. Using a constant and non-null password value when loading a keystore via an input stream can introduce security risks. 

\subsubsection{Manual investigation}
Our manual analysis of the posts in the sample dataset indicated that 82\%, \ie 327 posts, are afflicted with at least one security risk, and only 17 posts had an accepted answer with a secure solution.\footnote{The term ``secure solution'' refers to either a warning that highlights the risk or a code snippet that patches the issue in an accepted answer.}
In particular, we found a total of 559 security violations out of which 368 violations were present in 209 posts with an accepted answer (\ie solved posts).
Table~\ref{table:SecRulesviolAA} presents an overview of these risks, which we discuss below.

\emph{R-01.}
There were a total of 34 posts that adopted a weak algorithm.
Out of these posts, 14 of them were solved posts, but none of them offered a secure solution.

\emph{R-02.} 
We identified 265 posts that used an insecure mode for encryption, 192 of them were solved posts.
There were 127 posts that employed the ECB encryption mode. Regrettably, among the solved posts, only 7 provided secure solutions.
We also found 164 posts utilizing CBC encryption mode (R-02-b), which is considered a bad practice for client-server scenarios due to its lack of integrity protection. Unfortunately, only 8 of the solved posts offered secure solutions, considering integrity.

\emph{R-03.}
We observed static keys in 100 posts. 
Merely 6 solved posts presented secure solutions.
There were 20 posts that used static salts for key generation from passwords.
We found 4 posts with fewer iterations than 1000, and 25 posts that utilized hard-coded passwords. 
Only 4 solved posts provided secure solutions.
Finally, in 33 posts, developers employed weak algorithms for key generation, and among solved posts, only 8 accepted answers offered a secure solution.

\emph{R-04.}
We found static initialization vectors (IV) in 48 posts. 
Among the solved posts, only 5 of them suggested a secure solution. 
We also found 1 post with badly derived IV.

\emph{R-05.} We found only one violation instance of this rule.
 
\begin{table*}[]
 \centering
 
 \hspace{1cm}
 \caption {Insecure patterns and bad practices (\footnotesize{bad practices are shown in \textcolor{RedOrange}{orange color}, and manual steps are documented in \textcolor{ProcessBlue}{blue color}  preceded by an $\rightarrow$} symbol.)}
 \ttfamily
\begin{tabular}{p{1cm}p{15cm}}

\hline
\textbf{RuleID} & \multicolumn{1}{c}
{\textbf{RegEx} (PCRE2)} \\ [3pt] \hline
R-01 & Cipher\textbackslash{}.getInstance\textbackslash{}((``|\&quot)(DES|\textcolor{RedOrange}{DESede}|RC2|RC4|RC5|Blowfish|chacha20) \\ [3pt]
\rowcolor[HTML]{EFEFEF}R-02-a & Cipher\textbackslash{}.getInstance\textbackslash{}``(?:AES|DES|DESede|RC2|RC4|RC5|Blowfish|chacha20)(?:\textbackslash{}/ECB)?''|Cipher\textbackslash{}.getInstance\textbackslash{}
(\&quot;(?:AES|DES|DESede|RC2|RC4|RC5|Blowfish|chacha20)(?:\textbackslash{}/ECB)?\&quot; \\
\textcolor{RedOrange}{R-02-b} & 
\textcolor{RedOrange}{Cipher\textbackslash{}.getInstance\textbackslash{}(``(?:AES|DES|DESede|RC2|RC4|RC5|Blowfish|chacha20)\textbackslash{}/CBC|Cipher\textbackslash{}.getInstance\textbackslash{}(\&quot;
(?:AES|DES|DESede|RC2|RC4|RC5|Blowfish|chacha20)\textbackslash{}/CBC}  \\ 
\rowcolor[HTML]{EFEFEF}R-03-a & (?:key\textbackslash{}s* = |secret\textbackslash{}s*\=   )(?:"|\&quot;|\{|new byte\textbackslash{}[\textbackslash{}]\textbackslash\{)|SecretKeySpec\textbackslash{}(`` \\[3pt]
R-03-b &  
PBEKeySpec\textbackslash{}(\textbackslash{}s*\textbackslash{}S+\textbackslash{}s*,\textbackslash{}s*"\textbackslash{}S+"\textbackslash{}s*,\textbackslash{}s*\textbackslash{}S+\textbackslash{}s*|PBEParameterSpec\textbackslash{}(``|salt\textbackslash{}s*\= (?:``|\&quot;|\{|new byte\textbackslash{}[\textbackslash{}]\textbackslash{}\{)
\\[3pt]
\rowcolor[HTML]{EFEFEF}R-03-c & (salt\textbackslash{}s*=|PBEKeySpec\textbackslash{}(|PBEParameterSpec\textbackslash{}() $\rightarrow$ \textcolor{ProcessBlue}{check the size of salt.} \\[3pt]
R-03-d &
PBEKeySpec\textbackslash{}(\textbackslash{}s*\textbackslash{}S+\textbackslash{}s*,\textbackslash{}s*\textbackslash{}S+\textbackslash{}s*,[1-9]\textbackslash{}d\{0,2\}|PBEParameterSpec\textbackslash{}(\textbackslash{}s*\textbackslash{}S+\textbackslash{}s*,[1-9]\textbackslash{}d\{0,2\}|salt\textbackslash{}s*=
 $\rightarrow$ \textcolor{ProcessBlue}{examine if the Iteration Count value is less than 1000.} \\[14pt]
\rowcolor[HTML]{EFEFEF}R-03-e & 
(?:password\textbackslash{}s*\= |pass\textbackslash{}s*\= )(?:``|\&quot;)|PBEKeySpec\textbackslash{}("
\\[3pt]
R-03-f & kgenerator\textbackslash{}.init\textbackslash{}(\textbackslash{}d+,\textbackslash{}s*\textbackslash{}w+\textbackslash{})  $\rightarrow$ \textcolor{ProcessBlue}{assess how the random value, \ie the second parameter, is generated.} \\[3pt]

\rowcolor[HTML]{EFEFEF}R-03-g &
SecretKeyFactory\textbackslash{}.getInstance\textbackslash{}((?:``PBEWithMD5AndDES''|\textcolor{RedOrange}{``PBKDF2WithHmacSHA1''})|MessageDigest\textbackslash{}
.getInstance\textbackslash{}(``SHA-1''\textbackslash{}N*\textbackslash{}n*\textbackslash{}N*\textbackslash{}n*\textbackslash{}N*.digest\textbackslash{}(Key
\\[14pt]
R-04 & (IV\textbackslash{}s*\=|InitVector\textbackslash{}s*\ =|InitializationVector\textbackslash{}s*\ =|IvParameterSpec\textbackslash{}()\textbackslash{}s*(?:``|\&quot;|\{|new   byte\textbackslash{}[\textbackslash{}]\textbackslash{}s*\{)\\[3pt]
\rowcolor[HTML]{EFEFEF}R-05 & KeyStore\textbackslash.load\textbackslash( 
  $\rightarrow$ \textcolor{ProcessBlue}{examine whether the password, \ie the second parameter, is hard-coded.}

\end{tabular}
\label{table:InsecurePatterns}
\end{table*}
 
\subsubsection{Large-scale analysis}
We collected code snippets associated with each security violation during our manual investigation.
These snippets helped us to develop a set of regular expressions, listed in Table~\ref{table:InsecurePatterns}, to locate security violations in the posts.
We conducted a large-scale investigation of security issues by applying these expressions to the posts in our full dataset.
These expressions could locate violations for rules R-01, R-02-a, R-02-b, and R-03-g, but manual inspection was necessary for R-03-c, R-03-d, R-03-f, and R-05.
For instance, R-03-d RegEx only filtered posts that might contain information about the iteration size for key generation, and we performed a subsequent manual check to identify the actual violations. 
We also inspected the results for R-03-a, R-03-b, R-03-e, and R-04-a that locate constant values.
This was to ensure that the identified constants were intended for use within the cipher object. 
For instance, when encountering a hardcoded password, it is crucial to confirm that it is intended for use as a password-based key generation rather than being applicable to other scenarios.
Consider post ID 28519570 as an example, where the specified password has nothing to do with key generation.

\begin{table}[]
\centering
\hspace{1cm}
\caption {Security violations in the full dataset}
\begin{tabular}{llll}
\hline
 \textbf{Rule-ID}       &  \textbf{\#Solved Posts} &  \textbf{\#Pending Posts} &  \textbf{\#Total} \\ \hline
R-01   & 300           & 329            & 629    \\ 
\rowcolor[HTML]{EFEFEF} R-02-a & 717           & 760            & 1,477   \\ 
R-02-b & 804           & 747            & 1,551   \\ 
\rowcolor[HTML]{EFEFEF} R-03-a & 382           & 356            & 738    \\ 
R-03-b & 60            & 51             & 111     \\ 
\rowcolor[HTML]{EFEFEF} R-03-c & 0             & 4              & 4      \\
R-03-d & 6             & 15             & 21     \\
\rowcolor[HTML]{EFEFEF} R-03-e & 90            & 92             & 182    \\ 
R-03-f & 1             & 3              & 4      \\
\rowcolor[HTML]{EFEFEF} R-03-g & 161           & 109             & 270    \\
R-04-a & 174            & 140             & 314    \\ 
\rowcolor[HTML]{EFEFEF}
R-05   & 2             & 2              & 4      \\ \hline
\end{tabular}
\label{table:SecRulesFullData}
\end{table}

We found a total of 5,305 issues in 3,174 posts, showing that \ie 92\% of posts suffer from on average 1.7 issues.
Table~\ref{table:SecRulesFullData} lists the distribution of violations in the full dataset.
These issues were mostly present in questions, but in 234 posts, 266 issues were introduced in the accepted answers.

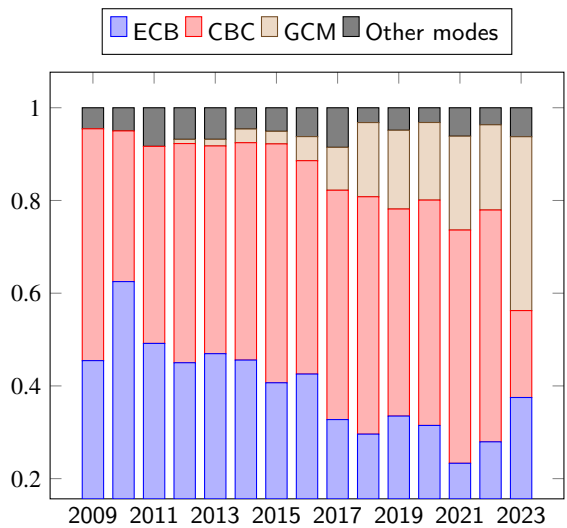
\begin{figure}
\centering    
\begin{tikzpicture}
\begin{axis}[
 width  = 0.48*\textwidth,
    ybar stacked,
	bar width=8pt,
    legend style={at={(0.5,1.15)},
      anchor=north,legend columns=-1},
    symbolic x coords={
2009,
2010,
2011,
2012,
2013,
2014,
2015,
2016,
2017,
2018,
2019,
2020,
2021,
2022,
2023
},
    ]
\addplot+[ybar] plot coordinates  {
(2009,0.454545454545455)
(2010,0.625)
(2011,0.491666666666667)
(2012,0.45)
(2013,0.469534050179211)
(2014,0.455737704918033)
(2015,0.406779661016949)
(2016,0.42560553633218)
(2017,0.327402135231317)
(2018,0.296)
(2019,0.33495145631068)
(2020,0.314741035856574)
(2021,0.233128834355828)
(2022,0.279411764705882)
(2023,0.375)
};

\addplot+[ybar] plot coordinates {
(2009,0.5)
(2010,0.325)
(2011,0.425)
(2012,0.472727272727273)
(2013,0.448028673835125)
(2014,0.468852459016393)
(2015,0.515254237288136)
(2016,0.460207612456747)
(2017,0.494661921708185)
(2018,0.512)
(2019,0.446601941747573)
(2020,0.48605577689243)
(2021,0.503067484662577)
(2022,0.5)
(2023,0.1875)

 };
\addplot+[ybar] plot coordinates {
(2009,0)
(2010,0)
(2011,0)
(2012,0.00909090909090909)
(2013,0.014336917562724)
(2014,0.0295081967213115)
(2015,0.0271186440677966)
(2016,0.0519031141868512)
(2017,0.0925266903914591)
(2018,0.16)
(2019,0.169902912621359)
(2020,0.167330677290837)
(2021,0.202453987730061)
(2022,0.183823529411765)
(2023,0.375)
 };
\addplot+[ybar] plot coordinates  {
(2009,0.0454545454545455)
(2010,0.05)
(2011,0.0833333333333333)
(2012,0.0681818181818182)
(2013,0.0681003584229391)
(2014,0.0459016393442623)
(2015,0.0508474576271186)
(2016,0.0622837370242214)
(2017,0.0854092526690391)
(2018,0.032)
(2019,0.0485436893203883)
(2020,0.0318725099601594)
(2021,0.0613496932515337)
(2022,0.0367647058823529)
(2023,0.0625)
 };
\legend{\strut ECB, \strut CBC, \strut GCM, \strut Other modes}
\end{axis}
\end{tikzpicture}
\caption{Share of encryption modes for AES}
\label{fig:distEncMode}
\end{figure}

We found 629 posts (300 solved) that used weak encryption algorithms.
Nevertheless, of 2,797 posts (1,406 solved) that utilized the recommended AES algorithm for symmetric encryption, the encryption modes were not secure in 2,425 posts (1,232 solved), weakening the advantage of AES.
In particular,
39\% were ECB and 48\% were CBC modes.
Figure~\ref{fig:distEncMode} illustrates the share of encryption modes in each year.
It seems that the adoption of ECB is decreased in recent years, whereas GCM is increased. Nevertheless, CBC has remained almost constant.

We found issues during the initialization of cipher keys in 998 posts (494 solved).
In particular, using static or constant keys was a common bad practice among programmers, with 738 posts containing this violation (382 solved). There were also 270 posts (161 solved) in which a secret key was generated from a password using insufficiently secure algorithms.
The ``PBKDF2WithHmacSHA1'' algorithm, which is a bad practice, was used in 197 posts (122 solved). ``PBEWithMD5AndDES'' which is insecure, was used in 11 posts (7 solved). Additionally, the hash function with ``SHA1'' which is insecure and the worst choice, was used in 60 posts (31 solved).

In 182 posts (90 solved) developers used hard-coded passwords. 
We found 314 posts (174 solved) that violated the rule for the initialization of cipher initialization vectors (IVs). 

\boxit[yellow!20]{
The number of security violations in the symmetric encryption posts on the StackOverflow website is significant, making it a misleading and dangerous information source, especially for novices.
}

\subsection{ChatGPT Analysis}
\label{subsec:RQ3}

With the advancement in generative AIs, developers are increasingly consulting them as a development assistant.
We examined whether ChatGPT can be relied upon from a security perspective to answer developer questions in 100 StackOverflow posts.

We considered a ChatGPT response as good as an accepted answer if it allowed us to easily reach the solution, similarly to the accepted answers on StackOverflow.
When we provided ChatGPT with the exact questions from StackOverflow posts.
We observed that 67 responses were relatively as good as accepted answers, but 33 responses were general and not to the point.
In terms of security, only three responses were free from violations. 
In particular, ChatGPT corrected 5 violations in 3 posts but introduced 4 new violations as well. 

We observed a notable improvement when for each question we explicitly asked ChatGPT to provide a ``secure'' solution. 
Particularly, 
ChatGPT provided secure solutions for 42 posts (58 remained problematic).

For each security violation that was still present in the ChatGPT answer, we explicitly asked ChatGPT to evaluate the security of the line associated with that violation. 
This resulted in the resolution of violations in 26 more posts.
Nevertheless, 32 posts were still problematic mostly because ChatGPT considered CBC as a secure encryption mode and even in an insecure scenario did not provide any warning.

\boxit[yellow!20]{
Security violations may exist in the code examples and hints provided by ChatGPT.
Users should exercise caution and follow a line-by-line security check to clear any doubts.
}

We investigated whether other prompts can help to identify violations without pinpointing the exact problematic line. 
Even the prompts such as ``provide a code example and make sure that every line is secure'' or ``examine your code example line by line and make sure that each line is secure'' or ``examine your code example and ensure that each line is secure for production'' did not help to address security issues and a manual line-by-line analysis is still superior.

We noted that when ChatGPT was prompted to provide a ``secure example for \textbf{production}'' the outcome did not change except providing warnings against hard-coded and constant values.

Even providing security guidelines in the prompt to ChatGPT did not result in a completely secure solution.  
For example, consider the following prompt for post ID 3451670:

\emph{
``[StackOverflow Question 3451670]
-> Please answer this question and provide a secure Java code example.
I expect a secure encryption algorithm, a secure encryption mode, and a secure Initialization Vector (IV).
Do not use hard-coded secrets, suggest best practices, and make sure that your example provides confidentiality and integrity.
Minimize the need for additional questions and ensure that every line in your code example follows this instruction.''}

Listing~\ref{lst:ChatGPTAnswer} presents the response.\footnote{https://chat.openai.com/share/8ce9eb6e-88b7-44dc-807a-51d04e7d446c}
ChatGPT identified the constant key, and it correctly suggested the secure ``GCM'' encryption mode as a best practice. However, violations such as ``static salt for key generation'' and ``hard-coded password'' were in the response without any proper warning. 

 \begin{lstlisting}[caption=Part of ChatGPT's answer to post ID 3451670, label=lst:ChatGPTAnswer]
public class SecureEncryptionExample {
    public static void main(String[] args) throws Exception {
        String username = "bob@google.org";
        String password = "Password1";
        String secretID = "BlahBlahBlah";
        String SALT2 = "deliciously salty";
        // Generate a secure random IV (Initialization Vector)
        SecureRandom secureRandom = new SecureRandom();
        byte[] iv = new byte[12];
        secureRandom.nextBytes(iv);
        // Derive a key from the password using a secure method
        byte[] key = deriveKey(SALT2, username, password);
        // Create a SecretKey object using the derived key
        SecretKey secretKey = new SecretKeySpec(key, "AES");
        // Instantiate the cipher with AES in GCM mode
        Cipher cipher = Cipher.getInstance("AES/GCM/NoPadding");
        ...
    }
    private static byte[] deriveKey(String salt, String username, String password) throws Exception {
        // Concatenate salt, username, and password
        String combined = salt + username + password;
        // Use a secure hash function (SHA-256) to derive a key
        MessageDigest md = MessageDigest.getInstance("SHA-256");
        byte[] hash = md.digest(combined.getBytes());
        return Arrays.copyOf(hash, 16);
    }
  } 
\end{lstlisting}

The suggested code, at line 23, included ``SHA-256'' hash function for key generation, which is not a best practice. Even worse, ChatGPT fails to address a critical issue with this construction in line 21. There is no domain separation between the three input fields. This means that both (``user1'', ``password'') and (``user'', ``1password'') result in the same derived key, potentially leading to catastrophic consequences. 
Anyway, 
when we asked about the security risk at line 6 (\ie static salt), ChatGPT responded: 
\emph{``Using a hard-coded string as a salt is generally not considered a good practice for security.
Ideally, a salt should be unique for each user and should be randomly generated for each password. Using a constant salt for all users could potentially make the system more vulnerable to certain types of attacks, such as rainbow table attacks''}. 
It then provided the code example in Listing~\ref{lst:ChatGPTAnswer2} as a replacement.

 \begin{lstlisting}[caption=Part of ChatGPT's answer to post ID 3451670, label=lst:ChatGPTAnswer2]
// Generate a secure random salt
SecureRandom secureRandom = new SecureRandom();
byte[] salt = new byte[16];
secureRandom.nextBytes(salt);
// Convert the salt to a Base64-encoded string for storage
String SALT2 = Base64.getEncoder().encodeToString(salt);

\end{lstlisting}

Subsequently, we asked, \emph{``Are there any other security risks in your provided code?''}, but it failed to detect hard-coded password. But in this step, it correctly suggested a key derivation function (KDF) instead of a hash function for generating the password-base key.
Upon a direct inquiry about the line where a hard-coded password existed, it offered a secure suggestion.

\section{Discussion}
\label{sec:Discussion}

We reflect on our findings about developers' challenges in symmetric encryption, security violations, and ChatGPT's support.

\subsection{Developer Challenges}
The most common root causes for developer issues were in key initialization (121), iv initialization (106), padding (89), encoding (54), and encryption mode (49).
Research suggests that the most significant reasons for these challenges among developers could stem from a lack of background knowledge~\citep{nadi2016, patnaik2019}, inadequate documentation~\citep{nadi2016, hazhirpasand2020, patnaik2019, acar2017}, usability issues~\citep{nadi2016,green2016}, low adoption of static analysis tools among developers~\citep{9680282}, and finally limitations in static analysis tools for detecting misuses~\citep{AutoDetecJavaCrypto,AfroseDetectionTools,b3}.

Security concerns were naturally predominant in the posts, and starting from 2020, we observed a consistent increase in concerns related to interoperability.
We were unable to locate any posts discussing compatibility concerns in the past two years (\ie 2022 and 2023), showing an improvement in this domain.

We came across 128 exceptions during the inspection of sample posts.
These exceptions, listed in Table~\ref{tab:SampExceptions}, initially appeared to be associated with padding and key-related issues, but further investigations revealed that the root causes may be different.
We found 61 cases where a \texttt{BadPaddingException} occurred. 
However, only 19 cases were directly related to padding problems. 
Most notably, 17 exceptions occurred during the encryption and decryption stages, primarily due to incorrect encoding.
We identified 14 cases with key-related issues, nine cases involving initialization vector (IV) problems, and two other problems. 
In 45 instances where an \texttt{InvalidKeyException} occurred, only 26 exceptions were directly linked to key initialization problems. 
Eight exceptions were thrown in the context of initialization vector (IV) issues, while the remaining 11 exceptions were associated with other problems.

\begin{table}[]
 \hspace{1cm}
 \caption {Top three recurrent exceptions}
\begin{tabular}{llp{3.5cm}}
\hline
 \textbf{Exception Type}                           &  \textbf{\#}                   &  \textbf{Exception Message (\#)}                          \\ \hline
\multirow{2}{*}{BadPaddingException}       & \multirow{2}{*}{61} & Given final block not   properly padded (36)  \\ 
                                           &                     &{\cellcolor[rgb]{0.902,0.902,0.906}} pad block corrupted (13)                      \\ \hline
\multirow{6}{*}{InvalidKeyException}       & \multirow{6}{*}{45}  & Illegal key size (14)                         \\  
                                           &                     & {\cellcolor[rgb]{0.902,0.902,0.906}}Invalid AES key length (9)                    \\  
                                           &                     & Key length not 128/192/256   bits (6)         \\ 
                                           &                     &{\cellcolor[rgb]{0.902,0.902,0.906}} Parameters missing (6)                        \\ 
                                           &                     & No installed provider supports this key (3) \\  
                                           &                     &{\cellcolor[rgb]{0.902,0.902,0.906}} Wrong Algorithm (3)                           \\ \hline
\multirow{2}{*}{IllegalBlockSizeException} & \multirow{2}{*}{19} & Input length must be   multiple of 16 (10)    \\  
                                           &                     &{\cellcolor[rgb]{0.902,0.902,0.906}} Last block incomplete in   decryption (3)     \\ \hline
\end{tabular}

\label{tab:SampExceptions}
\end{table}

\boxit[yellow!20]{
General exception messages are confusing as observed in 25\% of sample posts.
Precise and informative exceptions can help developers to identify root causes.
}

\subsection{Security Investigation}

Through our manual inspection of 400 posts, we found a total of 559 security violations in 327 posts.
From 251 posts with an accepted answer, only 32 posts were free from such violations.
In other words, from every 8 posts, only one is reliable.
Nevertheless, posts might contain short or incomplete code snippets requiring users to consult other posts which might introduce violations.
Even in 17 posts that included a direct question about security, only two received an accepted answer without any violation, and in two posts, new violations (R-03-e, R-04-a) were introduced in the accepted answer.

In our large-scale investigation of 3,426 posts, we found 5,305 issues in 3,174 posts.
We compared the distribution of security violations in the sample dataset versus the full dataset. 
We observed an increase, approximately two-fold, in the prevalence of weak algorithms in the full dataset. 
This could be attributed to a decline in the adoption of weak algorithms in recent years especially that our sample data was inclined toward recent and popular posts.
For the remaining rules, the distribution in the sample data closely mirrors that of the full dataset.

Several violations such as rules related to the presence of constant values may exist in the posts for demonstration purposes and providing working code examples (such as \texttt{salt=``1234''} or \texttt{key= ``passkey''}).
However, without a proper warning (which is almost absent in the posts), it is likely that novice programmers who resort to a copy-and-pasting approach, reuse these code examples, and at best, only substitute the constant values with new ones.
Even if we exclude so called demonstration mistakes,
for instance from 400 sample posts,
312 posts remain with 361 issues. 
By eliminating R-02-b (\ie the presence of CBC mode) these numbers were reduced to 180 posts and 197 issues, which are still problematic.

The minimum key length for AES should be 128 bits, but for applications demanding more security, NIST advises a 256-bit key~\citep{barker2020recommendation}.
Nevertheless, keys with a size of 128 were dominated for a long time. 
From 2013 onward, we observed a declining trend in 128-bit keys and an upward trajectory in 256-bit keys.
In 2018,
the distributions became equal.
This might be attributed to the release of JDK 9 that enabled developers to adopt stronger cryptographic algorithms by default, whereas in earlier versions they had to install ``JCE Unlimited Strength Jurisdiction Policy Files'' manually. It may also be driven by concerns over the quantum computing threat, which theoretically diminishes key sizes by half.
From 2019, a sharp decline in the use of 128-bit keys and a surge in 256-bit keys were evident. 
In addition, we observed an increase in the adoption of the more secure GCM mode in recent years, notably from 2018.
There were 
a total of 198 posts that employed GCM, and interestingly, only 30 posts (\ie 15\%) included violations.

Furthermore,
in 50 posts, we observed best practices such as ``PBKDF2WithHmacSHA256'' or ``PBKDF2WithH-macSHA512'' for generating a secret key from a password. 
However, the bad practice ``PBKDF2WithHmacSHA1'' was a more common practice, occurring in 197 posts. 
We also observed the adoption of insecure algorithms such as ``PBEWithMD5AndDES'' in 11 posts and ``SHA-1'' hash function for key generation in 60 posts.
There were 28 posts that employed ``SHA256'' and ``SHA512'' hash functions for key generation which is secure but not the best practice. 
We did not observe any additional noteworthy security best practices.

Finally, we looked at the profiles of users who contributed to accepted answers.
We did not observe a better performance of users with a crypto badge than those with a security badge. Likewise, reputation was not a good indicator of security. 
In particular, 
15 users who had a security-related badge (passwords, security, spring-security) contributed to 90 issues in 84 posts, 27 users with a cryptography-related badge (aes, aes-gcm,
cryptography, jce, encryption) made 199 issues in 158 posts, and 192 users with a reputation of at least 2000 made 410 issues in 336 posts.
This observation is inline with what previous work has shown about the success of developers in cryptography~\citep{8870184}.

\boxit[yellow!20]{
There are tons of StackOverflow posts that have security implications but are outdated. 
Generative AI assistants such as ChatGPT can be effective in flagging such content and improving developer awareness.}

\subsection{ChatGPT Support}

ChatGPT responses were correct, but in 33 cases tended to be overly general and missed the mark in addressing the question directly. Such responses may not be particularly helpful for beginners. A notable example highlighting this issue was when a developer posted a question (post ID 75266913) seeking help with an error in code generated by ChatGPT.
In the following, we discuss ChatGPT's performance through the lens of security.

ChatGPT correctly detected ``DES'' as an insecure encryption algorithm, However, it suggested ``3DES'' as a good choice although it is deprecated.
For instance, as we observed in response to post ID 22951606,\footnote{https://chat.openai.com/share/2d03613d-073e-4aaa-a644-9e2527c4dd75} it suggested to use ``3DES with a 192-bit key'', and included a note that \emph{``If you have the option, consider using AES with a 256-bit key for even stronger security''}.

ChatGPT correctly flagged ECB as an insecure encryption mode, but in some cases, \eg post ID 10759392,\footnote{https://chat.openai.com/share/c1150457-1c39-4019-94a1-152b93e53fad} it suggested to replace it with CBC, which could be problematic.
Indeed, 
While CBC itself is a secure encryption mode, it lacks integrity protection, making it insecure for client/server scenarios. However, ChatGPT may still recommend CBC as a secure choice, independent of its context. For instance, in response to post ID 18291987 concerning a client/server scenario, ChatGPT suggested CBC as a secure choice, despite its unsuitability in that context.\footnote{https://chat.openai.com/share/8149c107-659c-4995-a1f7-c9fe7d487a3c}
Nevertheless, in several cases, ChatGPT suggested using GCM instead of CBC for additional security.

ChatGPT did not always detect the use of constant values, even though we asked for a secure code example. 
It did catch this violation when we explicitly asked about the security of the exact line that included hard-coded secrets. 
Nevertheless, ChatGPT mostly suggests to replace the constant values, \eg ``with your own secrets'', which may still yield hard-coded secrets in the code by novices. 
For example, in answer to post ID 992019, ChatGPT suggested a hard-coded password but also included this warning: 
\emph{``You should use a strong passphrase, as a weak passphrase can be easily broken by an attacker. You should also make sure to use a secure method of storing the passphrase, as it is the key to your encrypted data.''}.
%

ChatGPT flagged ``SHA-1'' hash function for key derivation as insecure.
However, we observed instances where it suggested using the ``SHA-256'' hash function for key generation (e.g., post ID 992019)\footnote{https://chat.openai.com/share/9e4e0583-f5e4-4cce-84e7-8161da2a70ae}.
Although ``SHA-256'' is secure, best practices recommend to adopt a key derivation function, providing additional security measures like salting and iteration count to slow down the key derivation process, making it more resistant to brute-force attacks.
ChatGPT also correctly flagged ``PBEWithMD5AndDES'' as a non-secure password-based key generation technique, but it recommended ``PBKDF2WithHmacSHA1'' as a secure option, whereas it is not a good practice. In some cases, ChatGPT suggested ``PBKDF2WithHmacSHA256'' for higher security.

ChatGPT was not consistent in detecting violations of the number of iterations. For example, for post ID 4202499,\footnote{https://chat.openai.com/share/d5998eeb-7da5-416f-8b18-19865f7f5e71} it suggested to increase the iterations to at least 1000, whereas it missed the same violation in post ID 20888851.\footnote{https://chat.openai.com/share/9036dbd0-07dc-474a-8655-4c6e9c91437d}
Indeed, We realized that ChatGPT's response, even to the same question, might change over time. 
For example, when we queried about post ID 18291987,\footnote{https://chat.openai.com/share/571da441-3029-473d-8c0b-c26dec4a2be3} it correctly suggested to use GCM mode for the chat application.\label{chat}
However, when we posed the same question at a later time,\footnote{https://chat.openai.com/share/8149c107-659c-4995-a1f7-c9fe7d487a3c} it suggested to use CBC, which is certainly not a proper choice in a client-server scenario.

\boxit[yellow!20]{
ChatGPT is helpful but developers should know its limits.
It does not know every best practice and does not treat the same security violation consistently.
Therefore, ChatGPT cannot replace human expertise, and developers should remain alert throughout a ChatGPT session.
}

\section{Threats to Validity}
\label{sec:ThreatstoValidity}
In this section, we provide a summary of potential threats to the validity of this study, along with the strategies implemented to mitigate these threats~\citep{wohlin2012experimentation}.

\subsection{Threats to Internal Validity}

There are several threats to internal validity that may potentially affect the soundness of our study.

\emph{Data Collection:}
Our data collection approach for symmetric encryption posts might not be exhaustive. We primarily relied on the \texttt{Cipher.getInstance} pattern and did not investigate posts that may be relevant but lack the specified code snippet.
To expand our coverage, we also searched for posts tagged with ``java'' and containing the term ``symmetric'' in the question, resulting in 346 additional posts. However, a manual review revealed that most of these posts were unrelated to JCA.
Furthermore, we initially excluded 73 questions with ``java'' and ``symmetric'' tags, but further examination suggested that they might contain suitable code snippets for security analysis. Additionally, 228 posts contained the ``Cipher.getInstance'' pattern in answers instead of questions and lacked accepted answers.
Despite these challenges, we believe our data collection is comprehensive. While we may have missed a few posts with pertinent code snippets, our dataset largely aligns with our research objectives.

The inclusion of posts without accepted answer enabled a more comprehensive view of developer challenges including the ones that are still open (pending posts).
Nevertheless, security issues in solved posts have a higher impact than issues in pending posts. 
Therefore,  
we presented results for pending and solved posts separately.

\emph{Categorization Step:}
Human bias and errors are likely in manual investigations. Nevertheless, we employed a review strategy where each post was checked by at least two people and utilized a majority voting to clear disagreements.

\emph{Security Investigation:}
During the manual investigation, we may have missed some security risks due to human errors. To mitigate this, we implemented a double-checking process and used predefined rules from our base papers. It's possible that other security risks exist that we may have missed, but we strived to gather all relevant rules based on existing research.\\
During a large-scale study focusing on rules related to constant value violation, there is a possibility that our regular expressions might not have been able to identify constant or hard-coded values that were passed through variables. To address this concern, we implemented a strategy that involved using regular expressions to limit the number of posts we processed automatically, while manually reviewing the remaining posts to minimize the likelihood of missing any security issues.

Additionally, it is important to acknowledge that our collection of regular expressions may not cover all possible cases comprehensively. For instance, there is a chance that we might overlook insecure code snippets related to the constant key, as we primarily considered keywords such as \texttt{key} and \texttt{secret} appearing at the end of variable names. However, it's worth noting that, during our manual analysis, we observed a high likelihood of users using variable names containing these keywords.

Furthermore, including a large number of keywords in our regular expressions can make the detection process more challenging. Therefore, while our approach has its limitations, we have taken steps to mitigate these issues and enhance the overall accuracy of our security assessment.

\emph{ChatGPT Analysis:}
In our security analysis of ChatGPT responses, we acknowledge that human error and bias can impact the results. To address this concern, we implemented a double-checking system to enhance the accuracy of our findings. Additional reviews were conducted to minimize inaccuracies and improve overall precision.

\subsection{Threats to External Validity}

Threats to external validity stem from considerations of the generalizability of our study's results. It is crucial to acknowledge that our empirical investigation concentrated specifically on Java-related queries on StackOverflow. Hence, the extent to which our findings can be applied to other programming languages or platforms may be constrained. Furthermore, we investigated security risks based on the presence of insecure patterns, which may not uncover all conceivable vulnerabilities.

In evaluating the efficacy of LLM-generated code in addressing cryptographic challenges and ensuring the security of the generated code, we utilized the GPT-3.5 model as a representative of widely used models (chosen for its availability and popularity). It is important to recognize that this choice may limit the generalizability of our results to other models. To address this limitation, we plan to explore other models in the future, facilitating a comparative analysis that can yield more comprehensive insights into the security, correctness, and completeness of AI-generated code


\section{Related Work}
\label{sec:relatedwork}

We discuss related work in three key areas: code snippets security, cryptography challenges and misuses, and finally, evaluation of AI-generated code.

\subsection{StackOverflow code snippets security}
Several studies have examined the security of code snippets from StackOverflow. 

Verdi et al.~\citep{CppSecurity} revealed a concerning trend where a substantial number of insecure C++ code snippets had been transferred into GitHub projects.
Firouzi et al.~\citep{CSSecurity} studied C\# code snippets that include unmanaged code and found 67 code snippets with dangerous functions that
can introduce vulnerability (e.g., buffer
overflow) if not used with caution.

André et al.~\citep{Andre2022} investigated the security issues that developers encounter in WebAssembly, and they found that the topmost issues attributed to authentication in Blazor WebAssembly.

Meng et al.~\citep{JavaSecurity} examined code snippets for Java security. They observed an increasing trend in the adoption of third-party security frameworks like Spring Security for authentication and authorization. Also, they discovered that many accepted answers contained security flaws such as weak hash functions (MD5), SSL/TLS compromises, and disabled CSRF protection.

Rahman et al.~\citep{PythonSecurity} analyzed 529,054 Python code blocks from 44,966 answers posted on Stack Overflow. They identified 3,685 code blocks that exhibited at least one of the six critical insecure coding practices, namely code injection, cross-site scripting, use of insecure ciphers, insecure communications, race conditions, and insecure data serialization.

Fischer et al.~\citep{AndroidSecurity} employed machine learning techniques, training a model with 1,360 security-related code snippets sourced from StackOverflow answers. This model was subsequently applied to 3,834 code snippets, revealing that 30.28\% of them were identified as insecure.

\subsection{Cryptography challenges and misuses }
Several studies have explored the 
difficulties faced by developers in dealing with cryptography~\citep{AndroidSecureCode, LessonsCrypto, CryptographicMisuse, CryptoGuard}. 

Hazhirpasand et al.~\citep{9609232} clustered cryptography questions on StackOverflow, inspected a subset of them, and discussed why developers struggle in this domain.
They also analyzed JCA misuses in 489 open-source projects on GitHub and found that only 15\% of repositories were free from misuse, leaving 85\% susceptible to security issues~\citep{hazhirpasand2020}. 
The investigation of GitHub projects also revealed that approximately 64\% of cryptographic solutions in each project were not secure~\citep{8870184}.

Gajrani et al.~\citep{sPECTRA} revealed a troubling statistic indicating that a significant 90\% of applications available across diverse app stores are susceptible to exploitation due to cryptographic vulnerabilities.
Lazar et al.~\citep{2637237} conducted a systematic study of cryptographic vulnerabilities in practice.
Hazhispasand et al.~\citep{hazhirpasand2021_a} investigated cryptography vulnerability reports on the HackerOne bug bounty platform.

Several studies~\citep{piccolboni2021, kruger2017, zhang2022example} introduced tools for misuse detection. However, the adoption of static analysis tools is low among developers~\citep{9680282}.
To ease the adoption of static analysis tools for mainstream developers, 
Pourhashem et al.~\citep{Pourhashem2023} developed NASRA (NAturalistic Static pRogram Analysis), a framework that enables developers to define static program analyses in natural language, and they showcased how NASRA can be applied to uncover cryptography misuses in Java programs.
Nonetheless, existing tools are not able to catch every mistake. 
Braga et al.~\citep{8109084} found that coverage of static tools for finding cryptography missuses is far from good. Zhang et al.~\citep{AutoDetecJavaCrypto} conducted a comprehensive review of six crypto misuse detection tools by applying them to 200 Apache projects.
They found that there is no single tool being universally superior and that improvement in inter-procedural analysis and context awareness are necessary to effectively find misuses.
Ami et al.~\citep{b3} introduced a mutation testing framework to systematically assess the effectiveness of crypto-API misuse detectors.

Sharmin Afrose et al.~\citep{AfroseDetectionTools} developed detailed benchmarks and executed several vulnerability detection tools for the comparison of their effectiveness and reported their findings. 
Overlooked issues include difficulties in resolving parameter values, insecure initialization vectors, insecure random number generation, and insufficient key lengths in cryptographic key generation.
Finally, Kafader et al.~\citep{Kafader2021} developed FluentCrypto, an API designed to abstract away the low-level complexities inherent in utilizing the Node.js native cryptography API.

\subsection{Evaluation of AI-generated code}
The recent advancements in AI assistant tools have motivated researchers to examine AI-generated code. 
Fu et al.~\citep{fu2023security} analyzed 435 code snippets generated by Copilot in public GitHub projects and found that 35.8\% had Common Weakness Enumeration (CWE) issues across various programming languages.
Pearce et al.~\citep{pearce2022asleep} studied Copilot's performance in 
suggesting code related to 89 scenarios that were subject to MITRE's ``Top 25'' CWEs.
They discovered that around 40\% of the generated programs include vulnerabilities.
Asare et al.~\citep{asare2023copilot} conducted a study on GitHub's Copilot, examining its performance in various C++ and C scenarios. The findings revealed that while Copilot generated insecure code, it was not as bad as human developers in producing insecure code. Sandoval et al.~\citep{Sandoval} found that AI assistance in low-level C programming minimally affected security, introducing critical bugs only slightly more often (up to 10\%) than in cases without AI, indicating that the use of LLMs does not introduce new security risks.

Perry et al.~\citep{Perry_2023} studied how individuals utilize an AI code assistant, constructed using OpenAI’s Codex, to address security-related tasks across various programming languages. They found that participants who used the AI were more likely to introduce security vulnerabilities, yet they often perceived their insecure solutions as secure. Interestingly, those who invested more effort in crafting their queries to the AI tended to generate more secure solutions.

Kavian et al.~\citep{Kavian2024-lv} developed LLMSecGuard, an open-source framework designed to enhance code security through the integration of LLMs and static security code analyzers. 
It uses hints from static security analysis tools to guide LLMs in writing secure code.

\boxit[yellow!20]{
We presented the first in-depth investigation of symmetric encryption challenges and security risks.
In contrast to previous studies, we also investigated the support of generative AI (i.e., ChatGPT) to fix these issues.
}


\section{Conclusion}
\label{sec:Conclusion}

We conducted a thorough examination of StackOverflow posts related to symmetric encryption in Java.
We delved into developers’ challenges and concerns, discovering that they struggle significantly with key and IV generations, as well as padding. 
We observed that security is a top concern among developers. 
Nonetheless,
we examined each post from a security perspective and discovered that security violations related to symmetric encryption abound, making StackOverflow a risky information source, especially for novices. 
We evaluated ChatGPT’s support in resolving these issues and found that it can be very helpful only if proper interaction is in place.
Nevertheless, ChatGPT does not substitute human expertise, and developers should remain alert.

\bibliographystyle{cas-model2-names}

\bibliography{paper-main}

\end{document}